# Impact of strain on metamagnetic transitions in $Sm_{0.5}Sr_{0.5}MnO_3$ thin films

M. K. Srivastava,[1,2] A. Kaur,[2] K. K. Maurya,[1] V. P. S. Awana,[1] and H. K. Singh[1#]

[1]National Physical Laboratory (Council of Scientific and Industrial Research), Dr. K. S. Krishnan Marg, New Delhi-110012, India
[2]Department of Physics and Astrophysics, University of Delhi, Delhi-110007, India

**Abstract**

$Sm_{0.5}Sr_{0.5}MnO_3$ thin films were deposited by DC magnetron sputtering on $LaAlO_3$ (LAO) and $SrTiO_3$ (STO) substrates. The film on LAO, which is under compressive strain, undergoes paramagnetic–ferromagnetic (PM–FM) transition at $T_C$ ~116 K and shows insulator-metal transition (IMT) at $T_{IM}$ ~115 K. The film on STO is under tensile strain and has $T_C$ ~112 K; and shows IMT at $T_{IM}$ ~110 K. Around ~80 K the film on STO shows a broad peak in the resistivity that could be seen as the reentrant IMT due to appearance of a metamagnetic state, the presence of which is confirmed by the discontinuous irreversible jumps in the magnetic field dependent isothermal resistivity at T<85 K. These signatures of the metamagnetic state are not seen in the film on LAO. The occurrence and absence of metamagnetic state in films on STO and LAO, respectively has been explained in terms of the control of the competing FM metallic and antiferromagnetic–charge ordered insulating (AFM–COI) phases by the different strain states in these films.

# Corresponding author; Email: hks65@nplindia.org

Among the low bandwidth (W) manganites $Sm_{1-x}Sr_xMnO_3$ (0.45≤x≤0.5), because of its proximity to the strong electronic phase coexistence regime and the charge order/orbital order (CO/OO) instability, occupies unique status.[1-8] Over an appreciable compositional range (0.4<x≤0.50) it shows the most abrupt insulator metal transition (IMT) among the manganites.[1-8] The ground state of this compound is ferromagnetic metallic (FMM) in the compositional range 0.3<x≤0.52 that transforms to antiferromagnetic insulating (AFMI) state at x >0.52.[5-8] The charge ordered insulator (COI) appears at x=0.4 and the ordering temperature ($T_{CO}$) increases from ~140 to 205 K as x increases in the range 0.4≤x≤0.6. All the FMM ground states appearing in the range 0.3<x≤0.52 show the colossal magnetoresistance (CMR).[5,6] Around x≈0.45, a very sharp (first order) paramagnetic insulator (PMI) to the FMM transition, which is one of the prerequisites of a material having large magnetocaloric effect (MCE), is observed.[7,8] Near half doping, $Sm_{1-x}Sr_xMnO_3$ has an intrinsic tendency towards phase separation (PS) due to competing FMM and AFM/COI components, which often creates frustration that in turn causes evolution of a metamagnetic state. This also makes the composition-temperature (x–T) phase diagram intrinsically fragile with respect to external perturbations such as electromagnetic field, pressure, lattice strain provided by the substrate, etc.[1]

The strain caused by the difference in the lattice parameters of the substrate and the material is one of the external stimulus that is believed to play a key role in controlling the fraction and size of the ferroic phases such as the FMM and AFM/COI, which coexist due to the strong coupling between the spin, lattice and orbital degrees of freedom.[11-15] The tensile strain stretches the $MnO_6$ octahedra along the film plane (compression in the out–of–plane direction) and hence enhances the Jahn–Teller (JT) distortion that favors the AFM/COI. In contrast, the elongation of the $MnO_6$ in the out–of–plane direction by the compressive strain weakens the JT distortion and favours the FM–double exchange (DE), which enhances the FMM phase.[11,12] The strain is generally relieved by formation of defects and it has been shown that the creation of defects also favors the FMM at the cost of COI.[11,12] The magnitude of PS, i.e., fraction and dimensionality of the coexisting electronic phases is strongly dependent on band width (W), which is controlled by the average size of the A-site cations of the $ABO_3$ lattice. The degree of phase separation increases with decreasing W. Consequently, at small W even a relatively smaller strain is expected to have more pronounced effect on magnetotransport properties.

Extensive investigations have been carried out on $Sm_{1-x}Sr_xMnO_3$ (0.45≤x≤0.52) having poly- and single-crystalline bulk forms.[2-10] However, not much investigation has been reported[15-19] on polycrystalline or epitaxial thin films of $Sm_{1-x}Sr_xMnO_3$ and the literature is scarce. As reported earlier[14-17], one of the reasons for this seems to be the difficulty in the growth of thin films, due to the enhanced sensitivity of the electronic phases to the substrate induced strain in low W manganites. Nonetheless, recently the role of strain on the magnetotransport properties of $Sm_{1-x}Sr_xMnO_3$ (x≈ 0.45–0.47) has been investigated by Saber et al.[18] and Srivastava et al.[19] However, the evolution of metamagnetic state that arises due to the competing FMM and AFM–COI phases in the vicinity of half doping in $Sm_{1-x}Sr_xMnO_3$ has not been studied in thin films. In the present letter, we report our investigation on the effect of compressive and tensile nature of substrate induced strain on phase coexistence and the metamagnetic state in $Sm_{0.5}Sr_{0.5}MnO_3$ thin films. Our results establish that the nature of strain plays significant role in controlling the fraction of coexisting phases and hence metamagnetic state in $Sm_{0.5}Sr_{0.5}MnO_3$.

$Sm_{0.50}Sr_{0.50}MnO_3$ thin films (thick ~200 nm) were deposited by on-axis dc magnetron sputtering[19] of 2″ stoichiometric target on (001) oriented single-crystal $LaAlO_3$ (LAO) and $SrTiO_3$ (STO) substrates maintained at 800 °C at a dynamic pressure of 200 mtorr of Ar (80%) + $O_2$ (20%). The average in–plane (IP) and out–of–plane (OP) lattice parameters of the bulk target used for sputtering has been found to be a~b ≈ 3.8414 Å and c/2= 3.8419 Å, respectively. Since the IP lattice constants are larger than that of LAO (a= 3.798 Å) the strain will be compressive. The LAO in this case will provide a compressive strain $\varepsilon_C = -1.143$ %, while STO (a= 3.905 Å) will provide a tensile strain $\varepsilon_T = 1.629\%$. The larger magnitude of the tensile strain could have stronger impact on the microstructural as well as magnetic phases and their coexistence. The as grown films were found to be insulating and therefore, were annealed in flowing oxygen at 900 °C for 12 hrs. Shorter annealing time resulted in poor characteristics. Here we would like to mention that annealing at 900 °C has been used in case of various manganite thin films[18-21] and no interdiffusion at the substrate-film interface has been seen either in case of LAO or STO substrates. Annealing temperature higher than 900 °C have also been used in several studies.[17,20,21] The structure/microstructure was characterized by high resolution X-ray diffraction (PANalytical X'PERT PRO MRD, θ–2θ and ω–scans) and the cationic composition was studied by energy dispersive spectroscopy (EDS). Temperature and magnetic field dependent magnetization was measured by using a commercial SQUID magnetometer (MPMS–Quantum

Design) and the electrical transport under magnetic field was studied by standard four probe method using the 14T PPMS (Quantum Design).

The thickness of these films was estimated to about 200 nm. The film thickness was measured by a DELTEK profilometer (error ~ 10 %) on a step structured film deposited alongside the other films. In addition we also used atomic force microscope to measure the height of the step that gives the film thickness and the error in this case corresponds to the surface roughness of the films. The cationic composition estimated from the energy dispersive X–ray analysis (EDS) of the film on LAO (L50) is found to be Sm/Sr/Mn~0.50/0.50/0.99 and negligible spatial variation was observed in composition. In case of films on STO (S50), Sr presence in the substrate makes the analysis difficult but as all these films were prepared together, the cationic composition and homogeneity of the two films are expected to be identical. The occurrence of only (00ℓ) reflections in the θ–2θ scan (Fig. 1) confirms that the films have grown along the out-of-plane direction on both the substrates. The OP lattice parameters estimated from the XRD data are $c_{LAO}$ = 3.8462 Å and $c_{STO}$ = 3.8266 Å, respectively for films on LAO and STO. The larger (smaller) OP lattice constant of L50 (S50) than that of the bulk target ($c_{bulk}$=3.8419 Å) confirms the presence of compressive (tensile) strain. The changes in the OP lattice parameters of the films as compared to the bulk ($\Delta c = c_{bulk} - c_{film}$) are $\Delta c_{LAO}$=–0.0043 Å and $\Delta c_{STO}$=0.0153 Å. This clearly shows that the strain has relaxed more in case of the film on LAO than on STO. The larger change in the OP lattice constant in film on STO is attributed to the larger tensile strain $\varepsilon_T$ = 1.629 %. We would like to mention here that at thickness ~200 nm, the strain is generally expected to be completely relaxed.[11,12] However, in the present case presence of strain could be attributed to the larger difference between the substrate and materials lattice parameters.

To evaluate the mosaicity, we carried out the ω-scan of both set of films. The rocking curve of (002) diffraction maximum of both films is shown in Fig. 2. The rocking curve of the L50 film is symmetric and the full width at half maximum (FWHM) is found to be ≈ 0.34º (inset of Fig. 2). This suggests that over all crystallinity of the film is very good. In contrast the rocking curve of the S50 film is (i) asymmetric, (ii) has larger FWHM and (iii) bears signatures of presence of another peak. The de-convolution of the rocking curve of S50 (main panel of Fig. 2) clearly shows the presence of two diffraction maxima. The occurrence of such features that reveal higher mosaicity in the film on STO could be attributed to the presence of strain inhomogeneity that is caused by differently strained film layer due to the variation in the degree of strain as a function of the film thickness. The film layers in proximity with the substrate are strained, whereas the layers above these are relaxed. This strain inhomogeneity is expected to be distributed throughout the film and is a function of the film thickness. To confirm that this behaviour is not caused by the substrate we also did ω-2θ scan (data not shown here) around the (002) diffraction maxima of the SSMO film and the substrates. In the ω-2θ scan the diffraction maxima of the film on STO shows asymmetry, whereas in case of the film on LAO do not exhibit this feature. We did not observe any asymmetry in the STO peak. The observed strain inhomogeneities in the tensile strained film on STO and absence of any such feature in the film on LAO could be attributed to the relatively larger difference between the IP lattice constant of $Sm_{0.5}Sr_{0.5}MnO_3$ bulk (3.8414 Å) and STO substrate (3.905 Å), viz., the larger magnitude of the tensile strain (1.629%) than the compressive strain (−1.143%). The results discussed above clearly establish that even at larger thickness ~200 nm, strain causes structural inhomogeneities.

The temperature dependent zero field cooled (ZFC) and field cooled (FC) magnetization ($M_{ZFC}$, $M_{FC}$) measured at $H_{dc}$ =100 Oe is plotted in the upper inset of Fig. 3. The paramagnetic – ferromagnetic (PM–FM) transition temperature ($T_C$) is 116 K and 112 K for L50 and S50, respectively. Further, as we lower the temperature $M_{ZFC}(T)$ shows a cusp like feature at $T_P \approx 62$ K (53 K) in L50 (S50), which shifts to lower temperatures in the $M_{FC}$. The ZFC–FC curves show strong bifurcation starting just below the $T_C$ and a sharp decline below $T_P$. This magnetization decline is relatively sharper in the film under tensile strain, viz., S50. The irreversibility of the ZFC/FC curves at T< $T_C$ and sharp drop in the $M_{ZFC}$ below $T_P$ has been recognized as the generic feature of cluster glass like state.[17,19,22] This feature is common in low W manganites and is caused by the competing AFM–COI clusters in the FM region.[5,17,19] The sharper drop in the ZFC-FC magnetization in S50 could be regarded as a proof of larger AFM–COI fraction.

The temperature dependent resistivity (ρ–T) measured at zero and 50 kOe applied magnetic field is plotted in Fig. 3. In the PM region, both the films have nearly similar temperature dependence but show drastically different behaviours at T<$T_C$. A sharp IMT, where the resistivity drops by three orders of magnitude, this transition occurs at $T_{IM}$ ~ 115 K and 110 K in films on LAO and STO, respectively. Application of magnetic field leads to sharp drop in the resistivity and IMT shifts to $T_{IM}$ (50 kOe) ≈168 K (≈160 K) for L50 (S50). The magnetic field induced average increment in the $T_{IM}$ is found to be 1.06 K/kOe (1.0 K/kOe) for L50 (S50) and is slightly larger than the values reported by Demko et al. for $Sm_{0.55}Sr_{0.45}MnO_3$.[23] Both the films show MR in excess of 99% over a wide temperature range around $T_C/T_{IM}$ (lower inset of Fig. 3). The broadening of the MR–T curve around $T_C/T_{IM}$ is suggestive of strong phase coexistence.

Deep into the FMM regime the temperature and field dependence of the resistivity of the two films are drastically different. The S50 film shows a distinct hump around T~80 K, which appears to be the reentrant IMT. Application of a magnetic field of 50 kOe appreciably suppresses this reentrant IMT suggesting that the transition could be related to the enhanced AFM–COI fraction, which gets transformed into the FMM by the applied magnetic field. The MR–T curve of S50 shows a second peak in the MR–T around the reentrant IMT (lower inset, Fig. 3). The fact that the remanent hump in ρ–T curve of S50 taken at 50 kOe remains at T ~80 K also supports the AFM–COI origin. The absence of such a feature in L50 suggests that the reentrant IMT could be related to the tensile strain that favors the AFM/COI phases.[11,12] Thus we attribute this reentrant IMT and the peak in the MR–T curve at T=80 K to the presence of a metamagnetic state due to the frustration caused by the enhanced AFM–COI fraction in the FMM regime.

For confirming the origin of the reentrant IMT and the nature of the metamagnetic state, that is, whether this feature has origin in enhanced AFM–COI fraction, we measured the magnetic field dependent isothermal resistivity (ρ–H (T)). The ρ–H (T) data of both sets of films are plotted in Fig. 4 (a, b). At all temperatures the ρ-H curves show irreversible behavior. At T= 130K, which is well above $T_C/T_{IM}$ the ρ–H curves of both films have nearly identical behavior, wherein the resistivity first decreases slowly up to H≈ 12 kOe (14 kOe) for L50 (S50) and then magnetic field induced sharp drop in resistivity occurs. In case of L50 the resistivity drops by more than two orders of magnitude, while in S50 the decrease is slightly smaller. The MR exceeds 99% at H=32 kOe in L50 and at H=39 kOe in S50. Just below $T_C/T_{IM}$, at T=105 K, the field induced drop in resistivity is sharper in S50, which also shows stronger irreversibility. However, none of the films show saturation tendency in MR up to H=50 kOe used in the present study. At further lower temperature, the field induced resistivity decrement as well as the irreversibility is reduced in the compressively strained film L50. The ρ–H(T) data of S50, which is under inhomogeneous tensile strain shows contrasting features. At T= 80 K, the ρ–H(T) curve of this films shows discontinuities in form of sharp jumps that occur at different fields in the field increasing and decreasing cycles. For example, in ρ–H (80K) curve in the field increasing cycle the jump occurs at an upper critical field $H_{C1} \approx$ 16 kOe while in the field decreasing cycles it shifted to smaller lower critical field $H_{C2} \approx$ 7 kOe. At T=25 K, the field effect is further reduced and the $H_{C1} \approx$ 37 kOe and $H_{C2} \approx$ 16 kOe are significantly larger than that at 80 K. The occurrence of sharp jumps in the ρ–H(T) curve coupled with the reentrant IMT observed at T= 80K, confirms the presence of a metamagnetic state in S50, which is under tensile strain.

The results presented above show that the occurrence of (i) reentrant IMT, (ii) second peak in the MR–T curve, (iii) sharp irreversible discontinuities in the ρ–H(T) of film on STO could be regarded as signature of metamagnetic state and are caused by the enhanced frustration due to increase in AFM–COI fraction in the FMM regime.[24,25] The AFM–COI fraction is enhanced by the tensile strain and the associated strain inhomogeneity in the film on STO. This is also supported by (i) the reduced sharpness of the resistivity curve, (ii) variation of $H_{C1}$ and $H_{C2}$ with temperature and decrease in field induced effect on resistivity. The absence of the metamagnetic state in compressively stained film on LAO clearly underlines the effect to be the consequence of the different strain states of the two films. The observed magnetic field induced decrease (increase) in the resistivity in the field increasing (decreasing) cycles believed to be caused by the transformation of the AFM–COI state in to FMM ones.[24,25] The strength (sharpness and irreversibility) of the metamagnetic (i.e., an AFM-FM) transitions at $H_{C1}$ and $H_{C2}$, as seen in film on STO (Fig. 4b) is controlled by the respective phase fractions. During the field increasing cycle the AFM–COI to FMM transformation causes the metamagnetic jump, while the recovery of the AFM–COI state in the reverse cycle results in a jump at lower critical field $H_{C2}$. The hysteretic nature of the ρ–H(T) curves throughout the temperature range is a signature of the first-order nature of this field-induced phase transition.[24] The hysteresis loop is not completely closed at low temperatures, even when the field is decreased to zero suggesting that fraction of the AFM-COI states could still be present at these temperatures.

In summary we have demonstrated that the tensile strain (e.g., in STO where ε= 1.63%) enhances AFM–COI phase fraction that causes metamagnetic state at $T<T_C/T_{IM}$, which undergoes a first order phase transition as confirmed by the hysteretic jumps in the ρ–H(T) curves at different critical magnetic fields in the field increasing and decreasing cycles. The absence of such jumps and significantly reduced irreversibility in the film under compressive strain (L50) is due to relatively smaller AFM–COI fraction. At temperatures above $T_{IM}/T_C$ the onset of the field induced phase transformations at H=12 kOe (H=14 kOe) in L50 (S50) shows the presence of AFM–COI clusters. The CMR, in excess of 99% in both films even at 130 K which is well above $T_C/T_{IM}$, is due to the magnetic field induced AFM/COI–FMM transformation. Thus our results confirm that the substrate induced strain plays a decisive role in determining the phase coexistence and metamagnetic state.

MKS is thankful to CSIR, New Delhi for a research fellowship. Authors at CSIR-NPL acknowledge the continued support from Prof. R. C. Budhani.

**Figure Captions**

Fig. 1: The θ–2θ scans of the $Sm_{0.5}Sr_{0.5}MnO_3$ thin films. The $Sm_{0.5}Sr_{0.5}MnO_3$ maxima are indexed and the adjacent higher intensity peaks belong to the substrate.

Fig. 2: The rocking curve (ω–scan) of the $Sm_{0.5}Sr_{0.5}MnO_3$ thin films. The main frame shows the measured ω–scan (solid circle), the two de-convoluted peaks ($P_1$-strain relaxed and $P_2$ strained) and the fitted ω–scan curve (fit) of the film on STO. The inset shows the ω-scan of the film on LAO.

Fig. 3: The upper inset shows the M–T plot of L50 and S50. The main frame shows the ρ–T data of the two films measured at 0 and 50 kOe magnetic field. The lower inset shows the MR–T data at H=50 kOe.

Fig. 4: The isothermal ρ–H data of the film on (a) LAO and (b) STO measured at different temperatures.

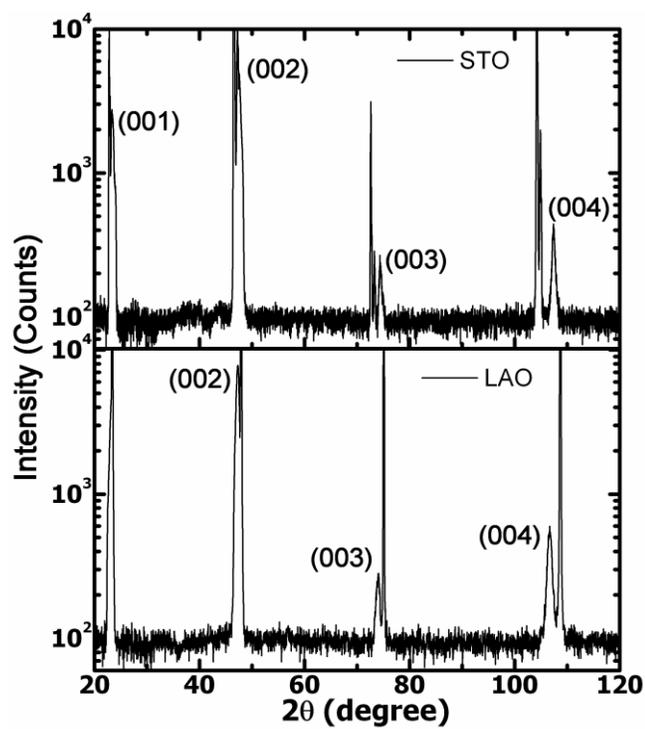

Fig. 1

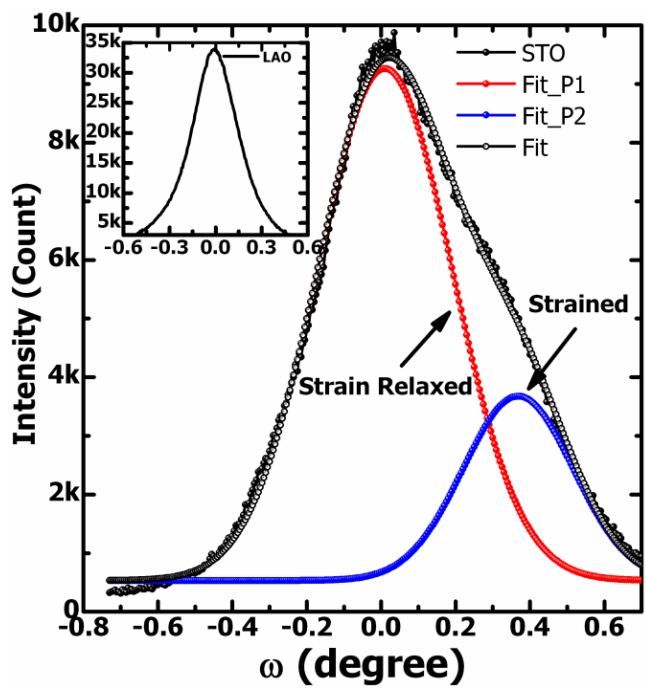

Fig. 2

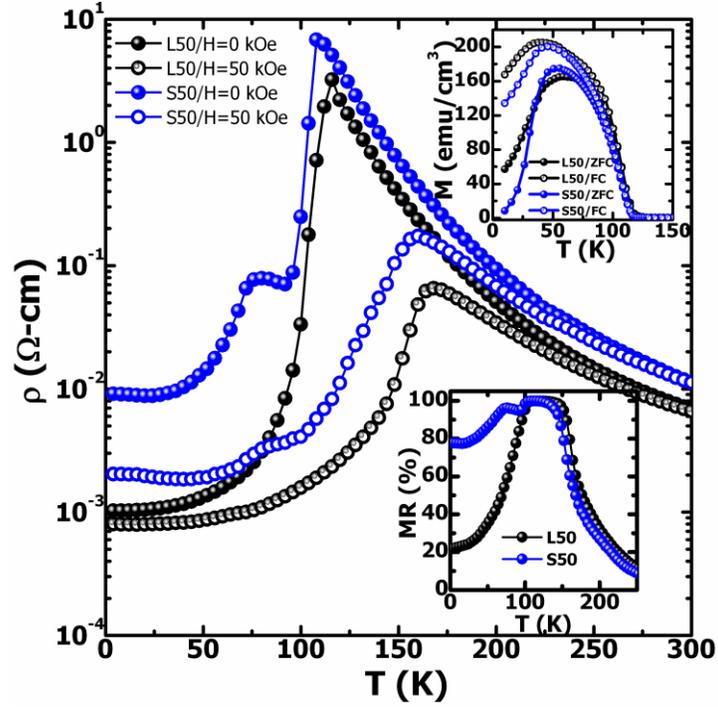

Fig. 3

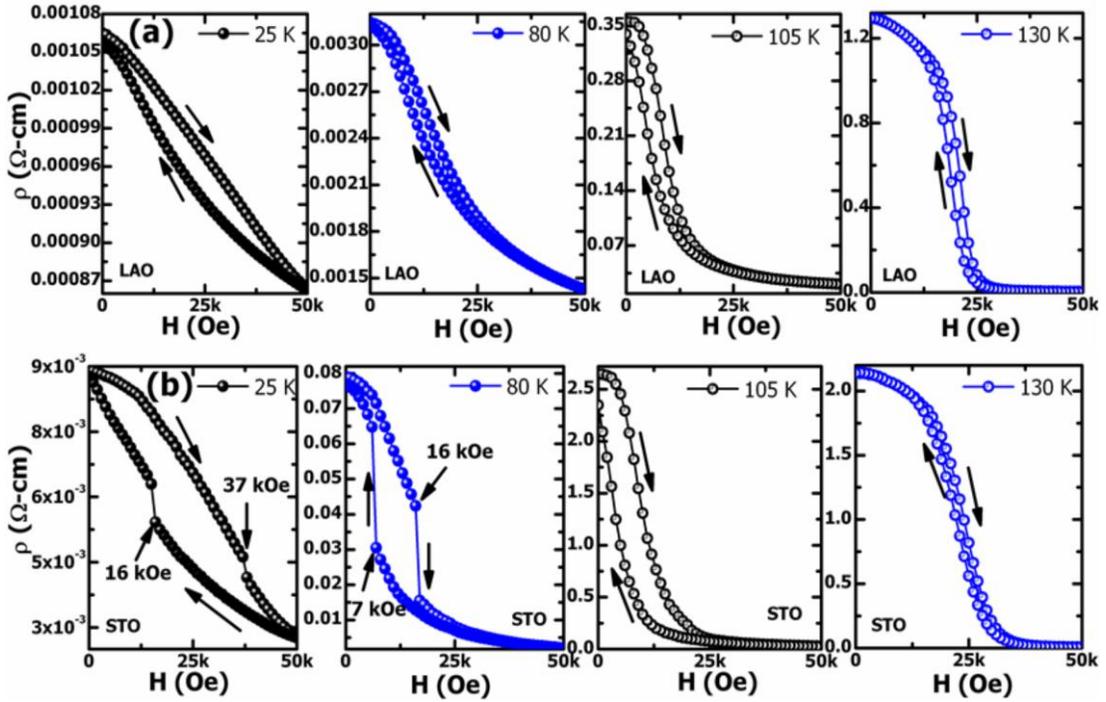

Fig. 4

13